\documentclass[a4paper,11pt]{article}
\usepackage{jinstpub} 
\usepackage{float}
\usepackage{lineno}
\usepackage{subcaption}
\usepackage{comment}

\proceeding{26$^{\text{th}}$ International Workshop on Radiation Imaging Detectors\\
Jul 6-10, 2025\\
Bratislava, Slovakia}

\title{New results from fast timing iLGAD sensor on Timepix4}

\author[a,b,1]{D.Oppenhuis,\note{Corresponding author.}} 
\author[]{on behalf of the Timepix4 Telescope group}
\affiliation[a]{Nikhef, Amsterdam, Netherlands}
\affiliation[b]{University of Amsterdam, Amsterdam, Netherlands}

\emailAdd{doppenhu@nikhef.nl}

\abstract{With the High-Luminosity Large Hadron Collider (HL-LHC) the number of collisions per bunch crossing increases. To cope with these high rates in the pixel trackers, per-pixel time measurements are required, which implies the need for fast sensors. The inverse Low-Gain Avalanche Detector (iLGAD) is one of the fast sensor options that is being investigated. This paper will show the results of an inverse Low-Gain Avalanche Detector (iLGAD) with a pitch of 55 \textmu m, a thickness of 250~\textmu m and a large-area (2~cm$^2$), bump bonded to a Timepix4 ASIC. Timepix4 has 195~ps time binning on each pixel and therefore an excellent ASIC to test the sensor. The sensor is characterised with radio-active source measurements in the lab, and during beam test at the CERN SPS North Area H8 beamline, where the Timepix4 telescope was used. The telescope has a time reference of 12~ps and a pointing resolution of 2.4 $\pm$ 0.1~\textmu m.\\
The iLGAD shows an almost uniform gain of approximately 4 and an efficiency of 99.6 $\pm$ 0.1\%. Without any corrections the obtained time resolution is about 750~ps. After timewalk and clock corrections the time resolution becomes 377 $\pm$ 7~ps. Grazing angle measurements have been done, which allow to measure the time resolution as function of depth of the charge deposition in the sensor. This provides more insight for the perpendicular time resolution.}


\arxivnumber{2509.09308} 

\notoc 
\begin{document}
\maketitle                
\flushbottom
\thispagestyle{empty}
\newpage
\setcounter{page}{1}

\section{Introduction}
\begin{figure}[h]  
\centering
\includegraphics[scale=0.28]{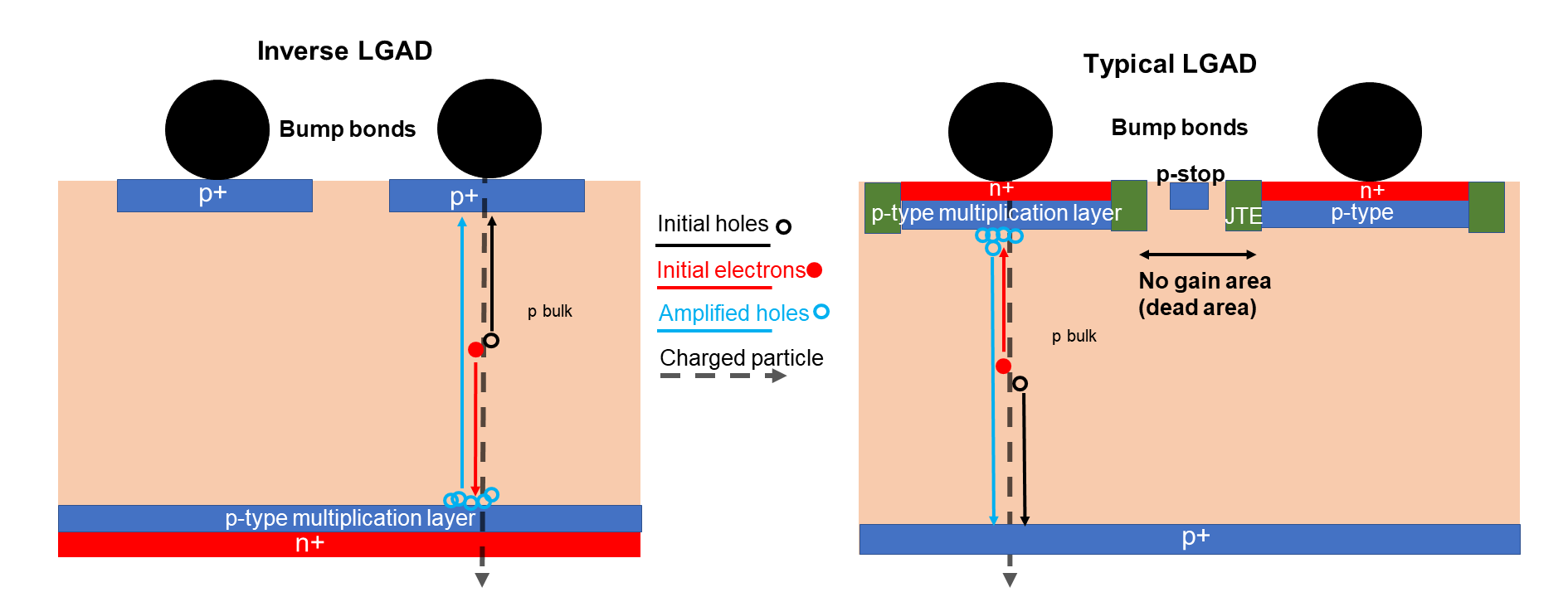}
\caption{A schematic view of two pixels of an iLGAD and a regular LGAD. The gain layer is on the backside for the iLGAD. The p+ pixel implant is connected via a bump bond to the ASIC. The traditional LGAD has an isolated area between the pixels where there is no-gain.}
\label{schem_ILGAD}
\end{figure}

With the upgrade to the High-Luminosity Large Hadron Collider (HL-LHC) the number of interactions per bunch crossing increases. In order to maintain the physics performance of the experiments at the LHC, time information must be used in addition to spatial information to distinguish all quasi simultaneous collisions per bunch crossing. This sets a requirement on the time resolution of the detectors of $\mathcal{O}(50~\,\text{ps})$ per hit. The currently best obtained time resolution with Timepix4 for a 100~\textmu m planar n-on-p type sensor with a pixel pitch of 55~\textmu m is $\mathcal{O}(150~\,\text{ps})$ \cite{timepix4telescope}. To reduce the timing error from Landau fluctuations the sensor should be thinner. This will decrease the total amount of signal, which will worsen the time resolution of the analog front end \cite{Heijhoff_2022}. Sensors that are thin, but still produce a large signal are so called Low-Gain Avalanche Detectors (LGADs). LGADs for Atlas HGTD and CMS MTD with a pitch of $\mathcal{O}(1~\,\text{mm})$ have achieved a time resolution of order $\mathcal{O}(30~\,\text{ps})$  for Minimum Ionising Particles (MIPs) \cite{sola2017ultra},\cite{giacomini2023lgad}. For vertex detectors, however, pixel pitches are typically $\mathcal{O}$(50-100~\textmu m).\\ 
An LGAD sensor has a gain layer, a region with a high electric field, where incoming electrons produce multiple electron-hole pairs via impact ionization in an avalanche process. The gain for LGADs is typically around 10-20 \cite{giacomini2023lgad}. In a traditional LGAD sensor design, the gain layer is implanted between the Silicon bulk and the pixel implant. The high field regions are separated via Junction Termination Extension (JTE + p-stop) structures (see figure \ref{schem_ILGAD}) or via trench isolation. The disadvantage of traditional LGADs is the no-gain distance of several tens of micrometers between pixels\footnote{For modern trench-isolated LGADs, the no-gain distance is limited to only a few micrometers.} \cite{giacomini2023lgad}, which lowers the fill factor $F=\frac{A_{gain}}{A_{total}}$. 
For a pixel pitch of 55 \textmu m and a no-gain region of 10~\textmu m the fill factor is only 67~\%. \\
A sensor design with a small pitch and a good fill factor is the inverse Low Gain Avalanche detector (iLGAD).
By placing the gain layer on the opposite side of the pixel electrodes (see figure \ref{schem_ILGAD}), iLGADs have a uniform gain layer, which has a positive effect on the fill factor. This results in gain over the whole sensor area. We research the timing performance of an iLGAD produced by Micron Semiconductor Ltd., which is bump bonded to a Timepix4 ASIC \cite{Llopart_2022}.

\section{iLGAD on Timepix4 ASIC}
An iLGAD hybrid sensor has been tested and characterised to investigate its charge and temporal performance. The sensor and readout ASIC are described here.


\subsection{iLGAD sensor under study}
The sensor has been produced by Micron Semiconductor Ltd. The inverse Low Gain Avalanche Detector (iLGAD) has 256 by 256 pixels with a pixel pitch of 55 \textmu m and a total sensitive area of 2~cm$^2$. The sensor has a 250 \textmu m thick p-type bulk and a highly doped p-type readout electrode, which is hole collecting. At the back there is a highly doped p-type layer for charge multiplication and the n-type implant (see figure \ref{schem_ILGAD}), forming the p-n junction. Hence the depletion starts from the back of the sensor and the pixels are depleted last. Therefore the sensor can only be operated fully depleted, otherwise the pixels are shorted together and will produce noise. The gain layer is depleted at \textasciitilde30~V and the full sensor at \textasciitilde160~V. The sensor is operated between 200 and 250~V, which gives a maximum over-depletion of 90~V. The outer ring of four pixels is designed without gain. These have been used to determine the gain. 
The iLGAD is bump bonded to Timepix4, which has a larger area of 448 $\times$ 512 pixels, and is connected to columns 0 to 255 and rows 8 to 263 of Timepix4.

\subsection{Timepix4 ASIC}
The Timepix4 is a hybrid pixel detector readout ASIC with a pixel pitch of 55 \textmu m \cite{Llopart_2022}. It can measure simultaneously the Time of Arrival (ToA) and Time over Threshold (ToT), which is proportional to the charge collected from the sensor. Time measurements are based on a  Voltage Controlled Oscillator (VCO). Every superpixel, a group of two by four pixels, has a VCO with a nominal frequency of 640 MHz. To further refine the ToA measurements, four phase shifted copies of the VCO are used, which results in a Time-to-Digital-Converter (TDC) bin size of 195~ps. A more detailed description of the Timepix4 ASIC is described in \cite{Llopart_2022}\cite{Heijhoff_2022}. The time resolution in hole collecting mode of the analog front-end and the TDC together is 122 ± 6 ps for a charge of 15~ke- \cite{Heijhoff_2022}.

\section{Laboratory test pulse and $^{241}$Am calibration}
\begin{figure}
     \centering
     \begin{subfigure}[b]{0.48\textwidth}
         \centering
         \includegraphics[width=\textwidth]{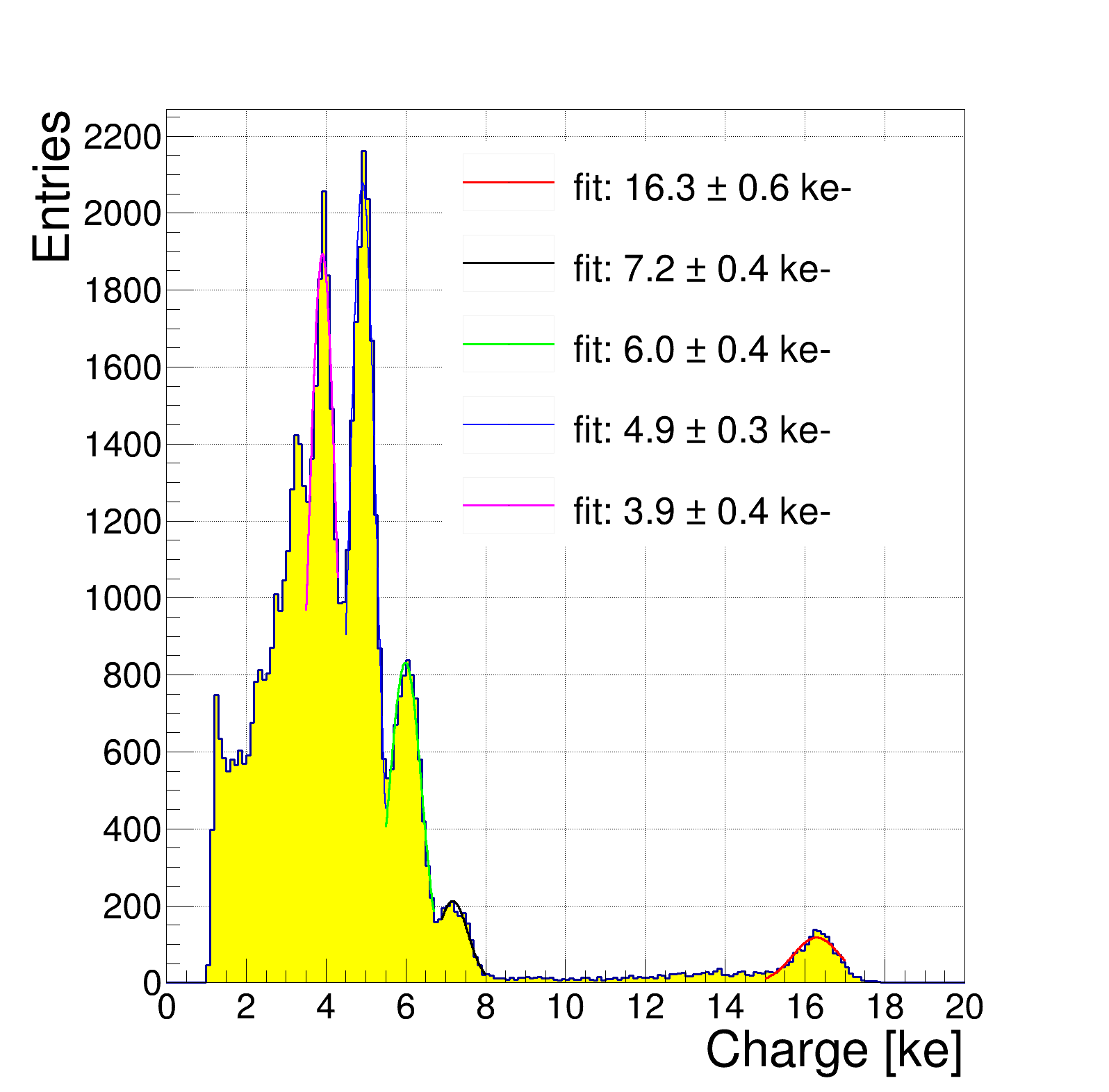}
         \caption{}
         \label{nogaincal}
     \end{subfigure}
     \hfill
     \begin{subfigure}[b]{0.48\textwidth}
         \centering
         \includegraphics[width=\textwidth]{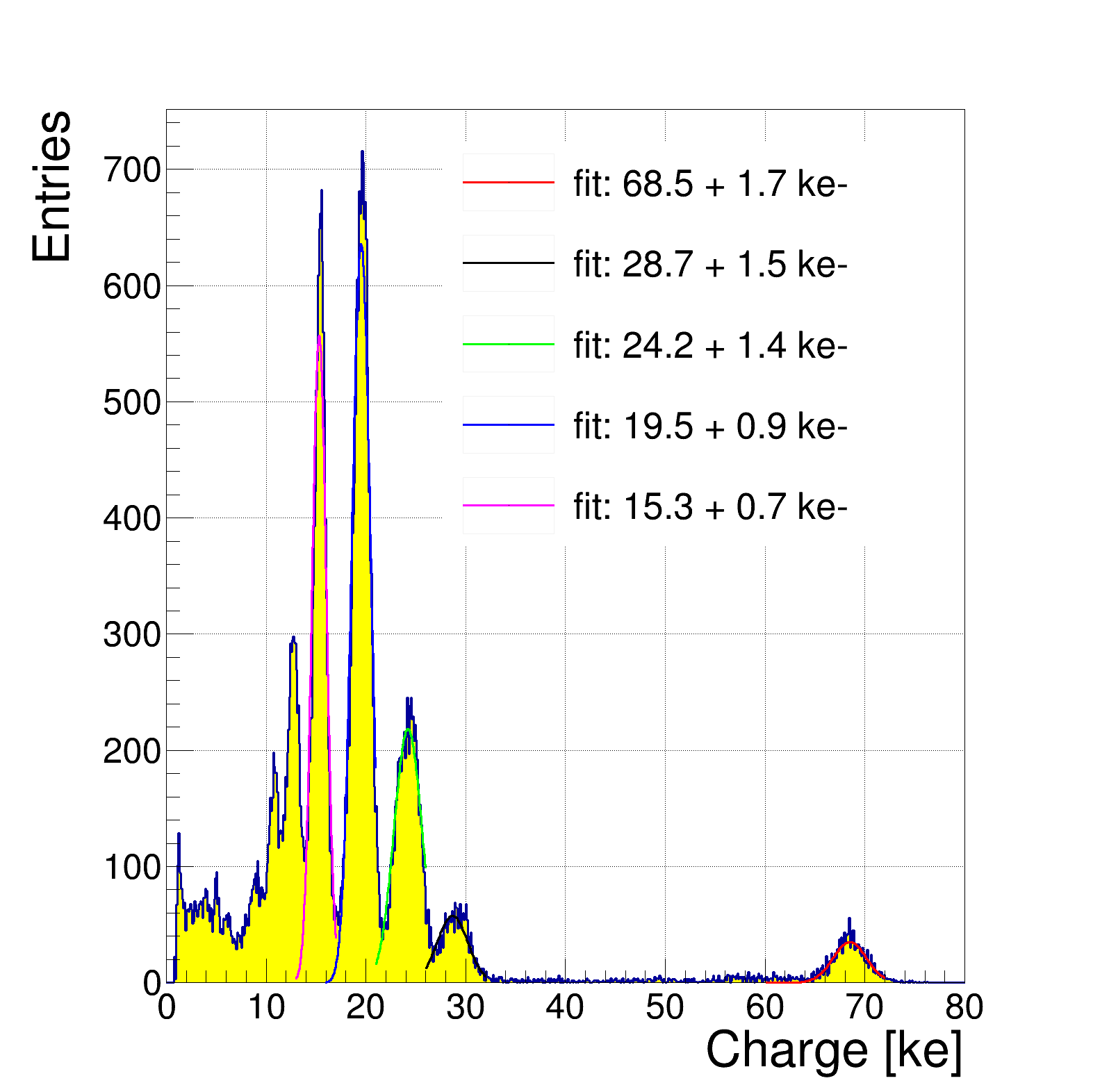}
         \caption{}
         \label{gaincal}
     \end{subfigure}
        \caption{Measured charge spectrum of the no-gain (left, col=0) and gain (right, col=245) pixels in a single column with an $^{241}$Am source. The deposited charge for a bias voltage of 200~V is shown. The mean value from a fit with a Gaussian function is listed in the plots. The uncertainty is the standard deviation of the fit.}
        \label{calibration}
\end{figure}

A set of laboratory measurements was performed as part of the sensor performance evaluation.
The gain of the sensor is measured with an $^{241}$Am source. Prior to the radioactive source measurements, a test pulse calibration has been performed to determine the (non-linear) ToT-to-charge calibration parameters per pixel \cite{masterthesis},\cite{delogu_charge}. $^{241}$Am and its daughter isotope $^{237}$Np emit X-rays with energies of 8.01, 13.9, 17.7, 20.7, 26.3 and 59.5 keV \cite{vicente2016caracterizaccao}. The average energy to create one electron-hole pair in silicon is 3.6~eV. Therefore peaks are expected at 2.22, 3.86, 4.92, 5.75, 7.3, 16.5 ke-, if the sensor would have no-gain.\\
The sensor has an outer ring of four pixels without a gain layer, these are used to check the test pulse calibration. 
In the energy spectrum of the no-gain area (see figure \ref{nogaincal}) there are no charge values observed above 18~ke-. 
In the fifth row and column these values above 18~ke- are present, but the mean is still lower than the gain area (row>5, col>5). This is attributed to charge sharing with the no-gain area and to fractional gain. 
Therefore the fifth row and column have been excluded from the analysis.
At a bias voltage of 200~V the $^{241}$Am source measurements shows that the charge calibration is accurate up to \textasciitilde1~$\%$ for the 16.5~ke- peak for the no-gain pixels, see figure \ref{nogaincal}. \\
The average measured charge in the gain area for the 59.5~keV Americium peak is 66.5~$\pm$~1.5~ke-, and 16.3~$\pm$~0.2~ke- in the no-gain area. 
This gives an average gain of $\textnormal{G}=(4.08\pm0.10)$\footnote{This claim is made with the assumption that the ToT measurement remains linear for large charges. The highest test pulse that can be injected via the test pulse calibration is around 20~ke-.}.

\section{Testbeam results}
An extensive test beam program has been performed at the CERN H8 beamline with a 180~GeV/c mixed hadron beam. The program is focussed on the temporal resolution of the iLGAD. First the efficiency is discussed and after that the time resolution at perpendicular and grazing angle beam incidence of the tested sensor.

\subsection{Timepix4 telescope}
To characterize the spatial and the time resolution of the sensor a testbeam campaign using the Timepix4 telescope has been performed. 
The Timepix4 telescope consists of eight planes, four 300~\textmu m for spatial resolution and four 100~\textmu m sensors for temporal resolution, see figure \ref{telschematic}. In the centre of the telescope a Device Under Test (DUT) is mounted on a stage, that can be rotated about the y- and the x-axis and moved in the x and y direction. The pointing resolution of the telescope at the position of the DUT is 2.4$\pm$0.1~\textmu m \cite{timepix4telescope}. Measurements have been performed with the tracks at perpendicular and grazing angle incidence. The temporal resolution of the telescope for a track is 92 $\pm$ 5~ps. For the iLGAD studied in this report the two microchannel plates (MCPs) downstream of the telescope are used as time-reference (see figure \ref{telschematic}). These MCPs have a combined temporal resolution of 12~ps \cite{timepix4telescope}. A detailed description of the telescope can be found in \cite{timepix4telescope}. 

\subsection{Hit efficiency}
\begin{figure}[t]
    \centering
    \begin{minipage}[t]{0.58\textwidth}
        \centering
        \includegraphics[width=\linewidth]{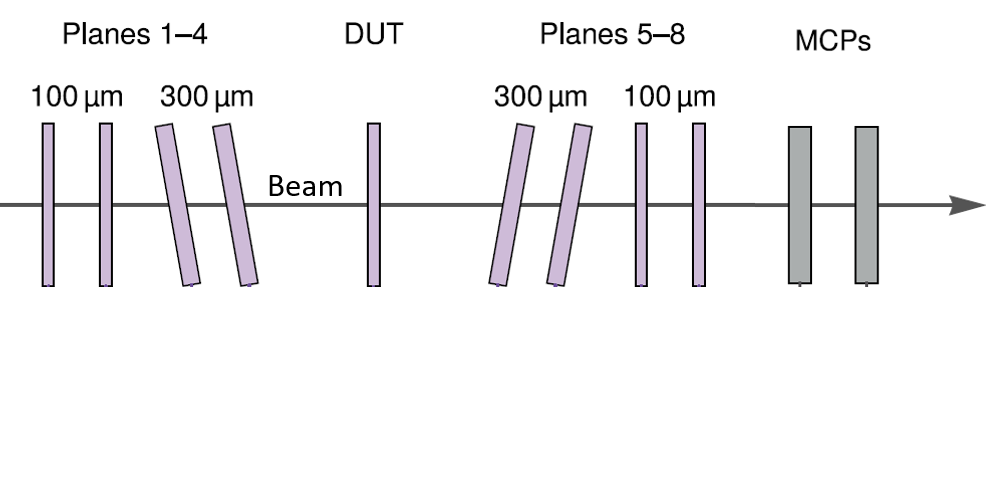}
        \caption{Schematic diagram of the eight telescope planes, the Device Under Test (DUT), and the two MCPs, adapted from \cite{timepix4telescope}.}
        \label{telschematic}
    \end{minipage}
    \hfill
    \begin{minipage}[t]{0.38\textwidth}
        \centering
        \includegraphics[width=\linewidth]{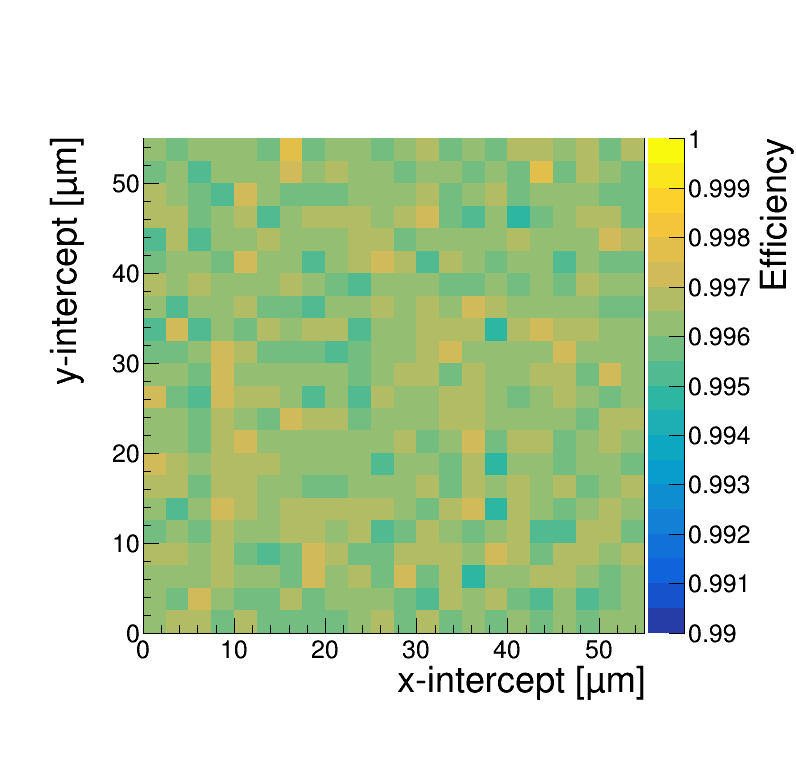}
        \caption{Intra-pixel efficiency of a 250~\textmu m thick iLGAD at 210~V and a threshold of 1000~e-. The average efficiency is 99.6\%.}
        \label{efficiency}
    \end{minipage}
\end{figure}

The track intercept position provided by the telescope is used to calculate the efficiency. The efficiency has been defined as the number of hits recorded by the DUT that are associated to tracks reconstructed with the telescope divided by the total number of tracks. One of the requirements of a track is that it has a cluster on all the eight telescope planes within a time window of 100~ns\footnote{A list of the standard selection requirements for the track reconstruction can be found here \cite{timepix4telescope}}. A cluster on the DUT is associated with a track when the cluster is within a time-window of 100~ns and at a distance of less than 1~mm in x and y direction with respect to the track. Tracks that pass the detector outside the sensor area, will not be taken into account for the efficiency calculation. Dead pixels in the sensor are taken into account, but their effect on the efficiency is negligible because they are outside the centre of the beam spot. The efficiency for a threshold of 1000~e- and bias voltages between 200-250~V has been measured at 99.6 $\pm$ 0.1\%. To calculate the intra-pixel efficiency all hits in a run are overlaid in one pixel. The efficiency is uniform over the whole pixel area, see figure \ref{efficiency}.

\subsection{Time resolution versus bias voltage and threshold} 
The time resolutions as function of bias voltage and threshold are investigated in the centre of the pixel matrix at col$\approx$130 and row$\approx$125. The size of the beam spot is \textasciitilde2 by \textasciitilde1.3~mm and the 300 pixels with most hits in the centre of the beamspot are used for the perpendicular time resolution.

\subsubsection*{Time resolution versus bias voltage}
\begin{figure}
     \centering
     \begin{subfigure}[b]{0.48\textwidth}
         \centering
         \includegraphics[width=\textwidth]{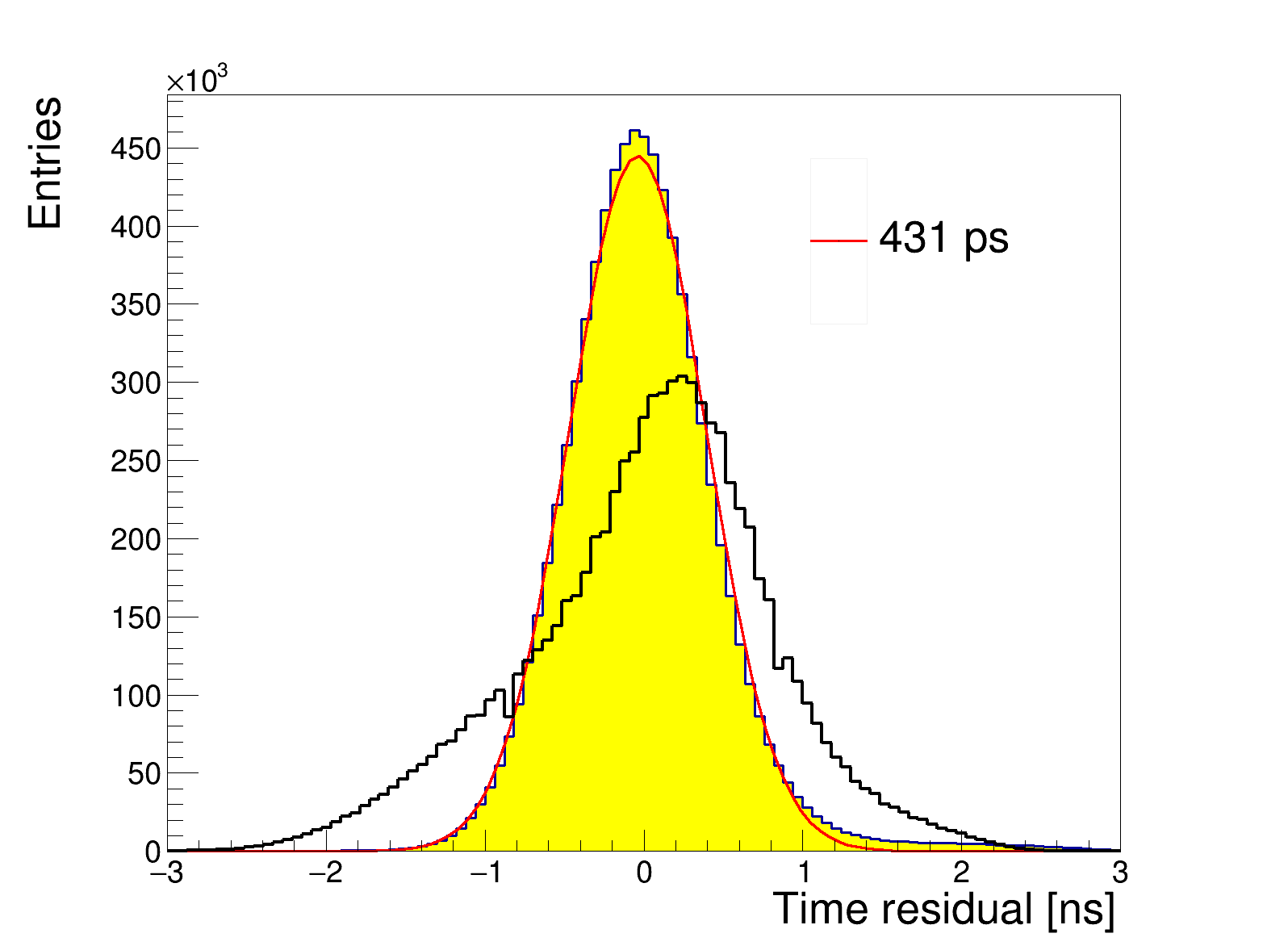}
         \caption{}
         \label{timeres250V}
     \end{subfigure}
     \hfill
     \begin{subfigure}[b]{0.48\textwidth}
         \centering
         \includegraphics[width=\textwidth]{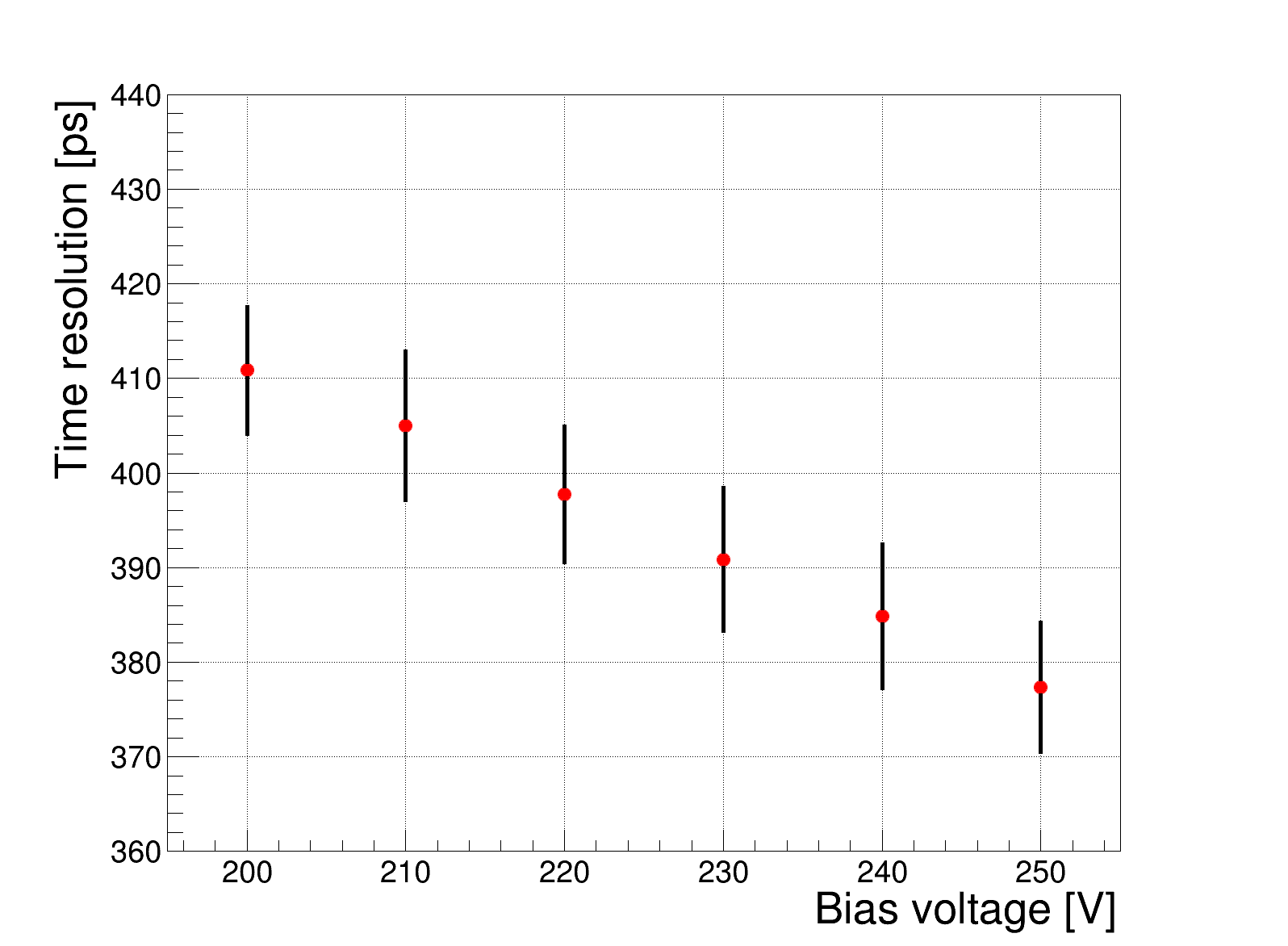}
         \caption{}
         \label{biasvstime}
     \end{subfigure}
        \caption{Time difference between MCP and Timepix4 hits for a bias voltage of 250V (left) and time resolution versus bias voltage (right) after per pixel correction. The black histogram in the left plot are the timestamps with a ToT cut. The yellow histogram is obtained after a global timewalk correction. The error bars in the plots show the spread in time resolution for different pixels}
        \label{time_resolution}
\end{figure}

A bias voltage scan is performed with a threshold of 1000~e-. By comparing the MCP timestamp with the Timepix4 timestamp, the time resolution can be determined. Because the MCP resolution is 12~ps the time difference distribution is dominated by the time resolution of the timepix4 measurement. The black line in Figure \ref{timeres250V} shows the difference between the MCP and the Timepix4 timestamp without any time corrections, but with a clusterToT cut of 700 $\times$ 25~ns, corresponding to about 50~ke-. The distribution is fitted via a chi-square fit with a Gaussian. The time resolution is defined as the standard deviation of the Gaussian fit. At a bias voltage of 200~V, with a ToT cut of 300 for the first pixel hit and a minimum cluster ToT of 700, the time resolution is 882~ps. An improvement for the time resolution to 746~ps is obtained by increasing the bias voltage to 250~V. This is lower than the 1.5~ns reported in \cite{svihra2024laboratory} for a sensor from the same batch, mounted on a Timepix3 ASIC and biased at 200~Volts. The difference can not be solely attributed to the difference in resolution of the TDC in the Timepix3 ASIC versus Timepix4 (\textasciitilde450~ps versus \textasciitilde60~ps).
 \\
 In order to improve the time resolution a timewalk correction is applied. Signals with a higher ToT will cross the threshold earlier (generally refered to as timewalk). Using the charge (ToT) of the signal it is possible to correct for this timewalk effect. The timewalk is modelled with the following empirical function:
 
\begin{align}
\Delta \textnormal{t}_{timewalk} = \frac{a}{q+b}+c,
\label{timewalk_form}
\end{align}
where $q$ is the charge deposited in the pixel, $a$ describes the curvature of the function, $b$ accounts for the charge offset and $c$ accounts for the time offset. Figure \ref{timeres250V} shows that the time resolution decreases by about 350~ps after using the same timewalk correction parameters (a,b,c) for all pixels, reaching 431~ps (467~ps) at 250~V (200~V).
\\
In the Timepix4 ASIC the same amount of charge can result in a more than 10\% difference in the measured ToT value due to pixel to pixel variations. Therefore, also a per pixel timewalk correction is determined. To reliably determine the parameters of the timewalk function, the 300 pixels that have most hits were selected. On average these pixels have a time resolution of 377~ps for a bias voltage of 250~V \footnote{The time resolution quoted here is also after VCO correction as described in \cite{timepix4telescope}}. In an earlier study using test pulses, it was found that the time resolution of the analog-front-end plus TDC is of order 120~ps \cite{Heijhoff_2022}. Hence the conclusion is that the iLGAD sensor dominates the time resolution. 
Figure \ref{biasvstime} shows that an increase in bias voltage to 250~V improves the time resolution, and could be improved with a further increase of the bias voltage. Increasing the bias voltage from 200 to 250~V gives an increase in gain of 2~\%, but the improvement in the time resolution is likely caused by more overdepletion of the bulk, giving a larger induced current. Since the depletion starts from the back, the electric field at the pixel electrodes is lower than in the rest of the sensor, while the weighting field peaks at the implant. The signal induction would be larger for a higher electric field close to the pixel electrode. The field around the electrode is estimated at a few kV/cm so the charge carriers are not yet velocity saturated.

\subsubsection*{Time resolution versus threshold}
The time resolution is also studied as function of threshold in the range [900-7000]~e- at a bias voltage of 250~V. As for the bias voltage scan, a per-pixel timewalk and VCO corrections are applied to achieve the best time resolution. Without corrections the time resolution deteriorates from 732~ps for a threshold of 900~e- to 1074~ps at 7000~e-. After a per-pixel timewalk correction the time resolution is almost constant at different thresholds (377~ps at 1000~e-, 391~ps at 7000~e-). For comparison, the time resolution of a 300 \textmu m p-on-n sensor is 387~ps at a threshold of 1000~e- and a bias voltage of 200~V. The depletion voltage of the 300~\textmu m sensor is $\approx$ 40~V, and hence the average E-field is somewhat higher than the iLGAD.

\subsection{Discussion of perpendicular results}
The perpendicular time resolution of the iLGAD is more than a factor two worse than for a planar 100~\textmu m sensor \cite{timepix4telescope}. To understand why the resolution is so much higher, it is important to consider the different contributions to the induced charge. When a particle traverses the sensor it creates electron-hole pairs. These initial charge carriers move owing to the applied bias voltage and induce a current according to the Schockley-Ramo Theorem \cite{ramo}. The weighting field for a sensor with a thickness larger than pixel pitch is peaked towards the pixel implant. The initial holes, which drift towards the pixel electrode, will induce a nearly constant current. From the initial electrons, which drift away from the pixel electrode, only the ones close the electrode will induce significant current because of the peaked weighting field. Electrons move faster than holes and therefore at first instance, the electrons will dominate the signal. The initial electrons that reach the back of the sensor will generate additional electron-hole pairs via avalanche multiplication in the high electric field of the gain layer. These secondary (amplified) holes will then drift towards the read-out electrode, but only will induce a significant current in the last part of their trajectory, again because of the highly peaked weighting field. Hence the current peak due to these secondary holes occurs later than the current owing to the initial electrons. The expectation is that the time resolution of 377 ps is mainly caused because the amplified signals arrive too late in the high weighting field region. The (integrated) signal already goes over threshold without the amplified signal. To understand this effect better it is interesting to disentangle the time resolution for different depths in the sensor with grazing angle measurements.

\subsection{Grazing angle results}
\begin{figure}  
\centering
\includegraphics[scale=0.25]{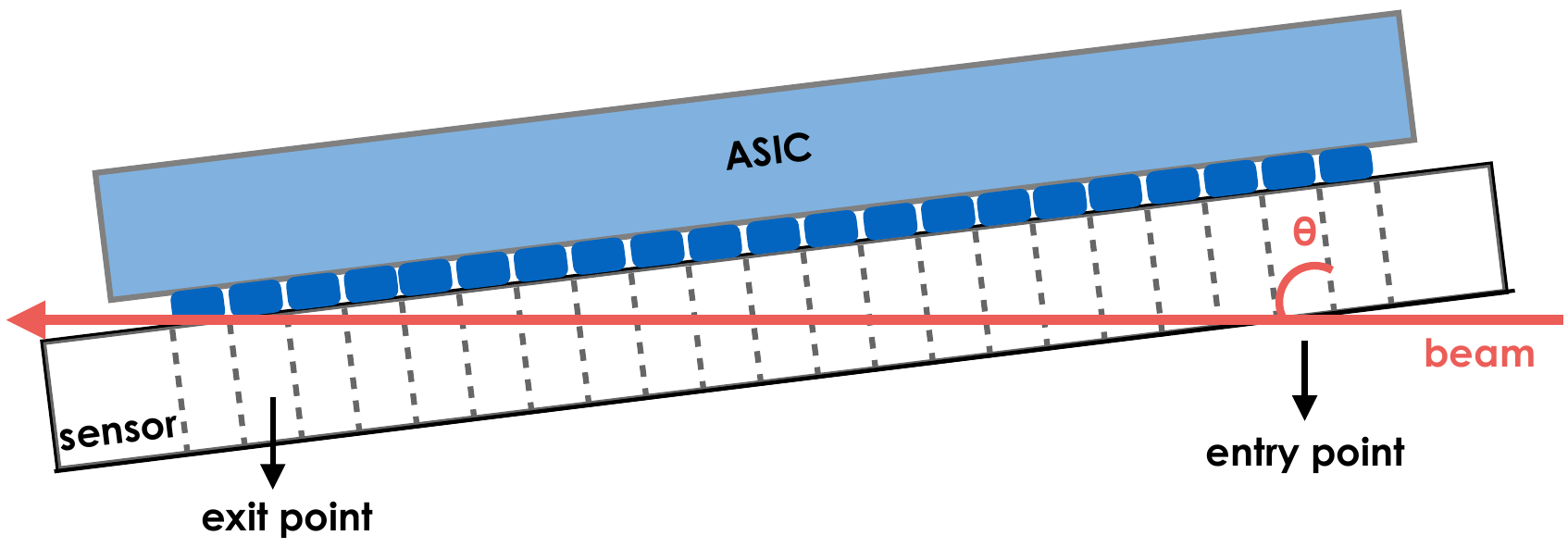}
\caption{Schematic view of the beam and the sensor in grazing angle setup. This figure is from \cite{Dall_Occo_2021}}
\label{schem_graz}
\end{figure}

In the so-called grazing angle method the particles traverse the sensor with tracks almost parallel to the sensor surface, see figure \ref{schem_graz}. Using this technique it is possible to study the time resolution as function of depth of the charge deposit in the sensor. A signal that is created close to the gain layer, will not show the double peak structure (initial holes and amplified holes) in the signal, because the amplified holes are created only a fraction in time behind the initial holes. \\
The tracks are parallel to the columns and in the direction of the row number. The tracks have a length of 100-150 pixels, depending on the angle. The entry point of the particles is at the gain layer and the exit point is at the read-out electrode.  Because the pixel pitch of the detector is 55 \textmu m, the expected charge deposit for the no-gain area is around 4000~e-, while for the gain area the expected signal is about 16000~e- per pixel. The time resolution will be investigated as a function of depth and threshold. 

\subsubsection*{Time resolution per depth versus threshold}
\begin{figure}[H]  
\centering
\includegraphics[scale=0.18]{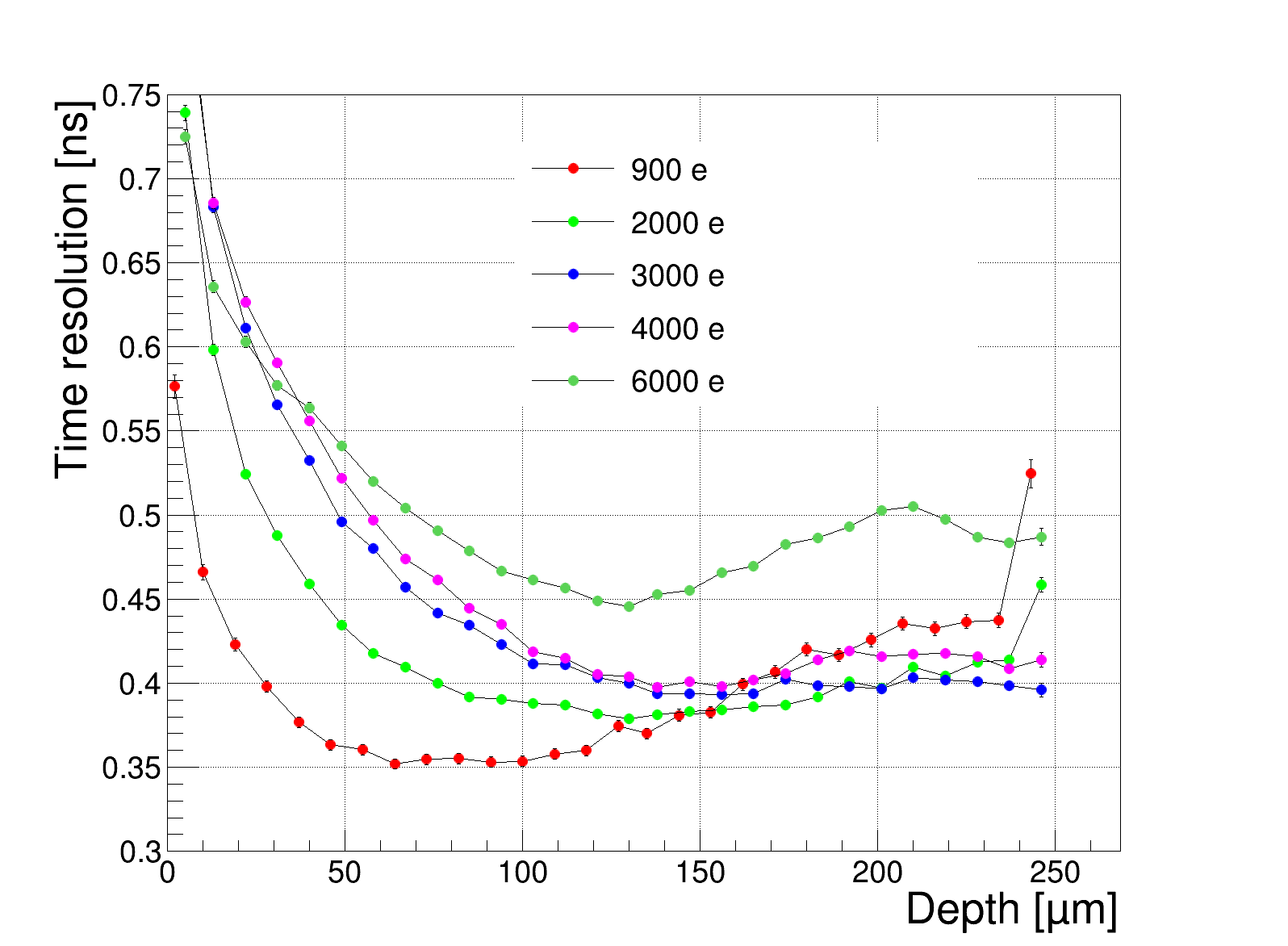}
\caption{The time resolution as function of depth for different thresholds. The time resolution is the best at a threshold of 900~e- at a depth of 75~\textmu m. The time resolution is determined with a Gaussian fit and the error bars represent the uncertainty of the fit parameter sigma.}
\label{timeresgain}
\end{figure}

The time resolution per depth is determined with the time difference between the recorded hits in a track and the MCP. The optimal time resolution in grazing angle mode has been obtained using a depth dependent timewalk correction. Close to the readout electrode the signal goes already over threshold on the initial charge carriers, while close to gain layer the amplified holes will also contribute to the signal shape. Because of this depth dependent signal shape, a timewalk correction per depth is needed. At a low threshold the signal goes over threshold on the initial holes and electrons. For a higher threshold above 4000~e- the signal will typically only go over threshold with the secondary holes coming from the gain layer. Therefore the timewalk parameters are also determined for different thresholds. 
 The best time resolution after timewalk correction is 359~ps for a threshold of 900~e- and at a depth of 75~\textmu m, see figure \ref{timeresgain}. The time resolution is obtained from a Gaussian fit and the error bars indicate the uncertainty of the fitted sigma parameter.\\
At the depth with the best time resolution the initial e-h pairs induce enough signal to cross the threshold. This initial e-h pairs induce a small current and the integrated signal has a low slew rate (shallow slope). Hence the threshold is crossed slowly, which results in significant jitter. The amplified signal will give a higher slew rate and hence if the threshold is raised such that it intersects this steeper slope, the jitter will be less. To avoid triggering on the initial signal, the threshold is raised from 900~e- to 7000~e-. However, we observe that the time resolution doesn't improve for higher thresholds. Only for signals coming from the back of the sensor the time resolution is somewhat better for a threshold of 2000, 3000 and 4000~e- compared to the 900~e- threshold, see figure \ref{timeresgain}. Simulations and (TPA) laser measurements should be done to understand the time resolution for different depths and thresholds.

\section{Conclusion}
A 250~\textmu m thick inverse LGAD with a 55 \textmu m pitch, bump bonded to Timepix4, has been characterised. The sensor has been tested with a 180 GeV/c charged hadron beam and has an efficiency of 99.6$\%$. At a bias voltage of 200~V an almost uniform gain is determined at $4.08\pm0.10$. The best perpendicular time resolution is 377~ps at a bias voltage of 250~V. The time resolution could be improved with higher bias voltages. The main contribution to the time resolution is from the sensor. \\
For grazing angle tracks we expected to see an improvement for charge generated close to the gain layer. The initial signal and the amplified signal are created at almost the same moment and therefore the time resolution should improve. However our measurements didn't show this improvement. To better understand these results, it would be helpful to carry out more extensive simulations. 
In the future there is a new production planned of thin 50~\textmu m iLGAD sensors from AIDAinnova. This thinner sensors will propably have a better time resolution. 

\section*{Acknowledgements}
This publication is part of the project FASTER with file number OCENW.XL21.XL21.076 of the research programme XL which is (partly) financed by the Dutch Research Council (NWO).
We would like to thank the University of Glasgow for the provision of the iLGAD sensor. 
This project has received funding from the European Union’s Horizon 2020 Research and Innovation programme under  GA no 101004761.




\bibliographystyle{JHEP}
\bibliography{biblio.bib}
\end{document}